\documentclass{aa}
\usepackage{graphicx}
\usepackage{txfonts}
\usepackage{hyperref}
\hypersetup{
  colorlinks=true,
  citecolor=blue,
  linkbordercolor={1 0 0},
  linkcolor=blue,
  urlcolor=blue,
  breaklinks=true
}

\usepackage{multirow}
\usepackage{longtable}
\usepackage[left,modulo,switch]{lineno} %switch option=>compile at least twice
\usepackage{array}
\usepackage{lipsum}
\usepackage{color}
\usepackage{float}

\newcommand{\rrlyr}{RR~Lyrae\ }

\newcommand{\rrly}{RR~Lyr\ }
\newcommand{\rrlyn}{RR~Lyr}
\newcommand{\ha}{H$\alpha$\ }

\newcommand{\kms}{~km\,s$^{-1}$\ }
\newcommand{\kmsn}{~km\,s$^{-1}$}
\newcommand{\cmsd}{~cm\,s$^{-2}$\ }
\newcommand{\cmsdn}{~cm\,s$^{-2}$}

\renewcommand{\aap}{A\textit{\&}A} % A&A bib
\begin{document}

\title{First observation of \ha redshifted emission in \rrlyn}

\subtitle{Evidence of a supersonic infalling motion of the atmosphere}

\author{
  D. Gillet\inst{1}
  \and B.~Mauclaire\inst{2}
  \and T.~Garrel\inst{3}
  \and T.~Lemoult\inst{4}
  \and Ph.~Mathias\inst{5}
  \and T.~de~France\inst{6}
  \and J-S.~Devaux\inst{7}
  \and H.~Boussier\inst{8}
  \and D.~Verilhac\inst{9}
  \and G.~Brabant\inst{10}
  \and J.~Desbordes\inst{11}
  \and O.~Garde\inst{12}
  \and the GRRR Collaboration\inst{13}\thanks{
  \textit{The \textbf{G}roupe de \textbf{R}echerche sur \textbf{RR} Lyr\ae} (GRRR) is an association of professionals and amateur astronomers leading high-resolution spectroscopic and photometric monitoring of complex phenomena such as the \rrlyr Blazhko effect.}
}

\institute{
  Observatoire de Haute-Provence -- CNRS/PYTHEAS/Université d'Aix-Marseille, 04870 Saint-Michel l'Observatoire, France \\ \email{denis.gillet@osupytheas.fr} %Gillet
  \and Observatoire du Val de l'Arc, 13530 Trets, France \\ \email{bma.ova@gmail.com} %Mauclaire
  \and Observatoire de Fontcaude, 34990 Juvignac, France %Garrel
  \and Observatoire de Chelles, 77500 Chelles, France %Lemoult
  \and Observatoire Midi-Pyr\'en\'ees, IRAP, Universit\'e de Toulouse, CNRS, UPS, CNES, Tarbes, France
  \and Observatoire des Tourterelles, 34140 Mèze, France %deFrance
  \and Observatoire OAV, 34290 Alignan-du-Vent, France %Devaux
  \and 84450 Saint Saturnin-les-Avignon, France %Boussier
  \and 26420 Saint Agnan en Vercors, France %Verhilac
  \and 26190 Saint-Laurent-en-Royans, France %Brabant
  \and 5, rue Edmond Gondinet, 75013 Paris, France %Desbordes
  \and Observatoire de la Tourbière, 38690 Chabons, France %Garde
  \and Observatoire de Haute-Provence, 04870 Saint-Michel l'Observatoire, France %GRRR
  %------------------------- pas de saut de ligne %
  \thanks{Based in part on observations made at the Observatoire de Haute Provence (CNRS), France}
}

\date{Received October 14, 2016; accepted August 22, 2017}

%= Abstract ==========================================================================%

% \abstract{}{}{}{}{} 
% 5 {} token are mandatory

%- 20160825 :
\abstract
  % context heading (optional)
  % {} leave it empty if necessary
   {The so-called \ha third emission occurs around pulsation phase $\varphi=0.30$. 
It has been observed for the first time in 2011 in some RR~Lyrae stars. 
The emission intensity is very weak, and its profile is a tiny persistent hump in the red side-line profile.}
  % aims heading (mandatory)
   {We report the first observation of the \ha third emission in \rrly itself (HD\,182989), the brightest \rrlyr star in the sky.}
  % methods heading (mandatory)
   {New spectra  were collected in $2013-2014$ with the \textsc{aurelie} spectrograph (resolving power $R=22\,700$, T152, Observatoire de Haute-Provence, France) 
and in $2016-2017$ with the \textsc{eShel} spectrograph ($R=11\,000$, T035, Observatoire de Chelles, France).
In addition, observations obtained in 1997 with the \textsc{elodie} spectrograph ($R=42\,000$, T193, Observatoire de Haute-Provence, France)
were reanalyzed.}
  % results heading (mandatory)
   {The \ha third emission is clearly detected in the pulsation phase interval $\varphi=0.188-0.407$, that is, during about 20\% of the period.
Its maximum flux with respect to the continuum is about 13\%. 
The presence of this third emission and its strength both seem to depend only marginally on the Blazhko phase. 
The physical origin of the emission is probably due to the infalling motion of the highest atmospheric layers, which compresses and heats the gas that is located immediately above the rising shock wave. 
The infalling velocity of the hot compressed region is supersonic, almost $50$\kmsn, while the shock velocity may be much lower in these pulsation phases.}
  % conclusions heading (optional), leave it empty if necessary 
   {When the \ha third emission appears, the shock is certainly no longer radiative because its intensity is not sufficient to produce a blueshifted emission component 
within the \ha profile. 
At phase $\varphi=0.40$, the shock wave is certainly close to its complete dissipation in the atmosphere.}


%= Title and authors ==========================================================================%

\maketitle

%
%=======================================================================================%
\section{Introduction}

The variability of RR\,Lyr, the brightest RR\,Lyrae star in the sky, has been discovered by the Scottish astronomer Williamina Fleming at Harvard \citep{pick1901}.
The light curve exhibits a period of about 0.5667 days (13.6\,h) that is attributed to pulsation.
It also presents amplitude and phase modulations with a period of about 39 days, the so-called well-known Blazhko effect \citep{bla1907}.
Its physical origin still remains a mystery, although several interesting hypotheses have recently been proposed to explain the Blazhko phenomenon \citep{smolec13, kovacs16}.
In addition, the pulsation and Blazhko periods change slightly during their respective cycles, and the ephemerides therefore have to be updated regularly \citep[e.g.,][]{leborgne14}.

\citet{preston65} reported the first detailed spectroscopic study of RR\,Lyr during a whole Blazhko cycle, but this investigation was limited to the rising part
of the pulsation light curve.
During this phase, \citet{preston65} observed line doubling in metallic lines as well as emission in the H$\alpha$ profile.
These phenomena were interpreted by means of a shock wave propagation, following the work of \citet{schwa52}, who showed that
a shock wave can produce these particular profiles in population \textsc{II} Cepheids.
\citet{preston65} also showed that during its outward propagation, the shock increases in intensity and velocity.
In addition, during the Blazhko cycle, the critical zone in the stellar atmosphere where the wave breaks moves up and down.
These observational results were modeled by \citet{fokin97}, who computed more than 20 nonlinear nonadiabatic pulsation models.
Although the code they used is purely radiative, convection is
thought to affect the results only weakly since the helium ionization zone is located deeper in the envelope
\citep{XCD98}.
In particular, \citet{fokin97} showed that five main shocks occur during a pulsation cycle.

In this paper, the pulsation phase is noted $\varphi,$ and $\varphi=0.0$ corresponds to the maximum luminosity.
The Blazhko phase is noted $\psi$ and has its maximum at $\psi=0.0$, the time of the highest luminosity amplitude observed during a Blazhko cycle.

\medskip

\begin{figure}[!h]
  \centering
  \includegraphics[width=\hsize]{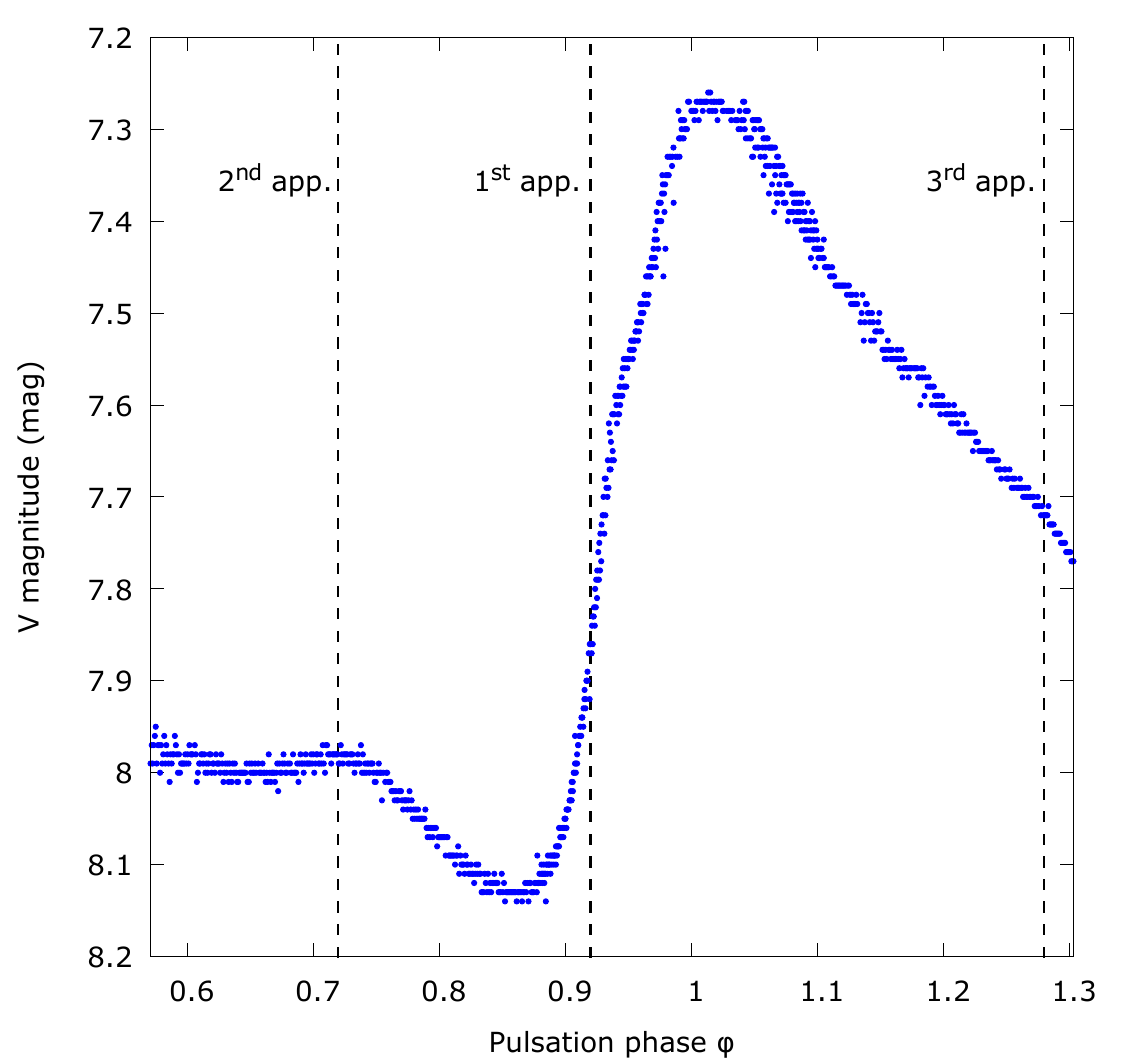}
  \caption{RR\,Lyr light curve covering the pulsation cycle from $\varphi=0.57$ to $\varphi=1.31$.
$V$ magnitudes were measured with the $G$ filter (490-580\,nm) corresponding to the $V$ photometric Johnson-Cousins bandwidth (490-590\,nm) on a Schmidt-Cassegrain 5'' telescope.
Observations were made between June 10 and 15, 2013, when the
Blazhko phase was in the interval $\psi=0.74-0.87$.
Dashed lines show the phases around which the three different emissions appear within the \ha profile.}
  \label{lumgraph}
\end{figure}

During a pulsation cycle in RR\,Lyrae stars, as we show in Fig.\,\ref{lumgraph}, there are three successive appearances of hydrogen emissions.
\citet{preston11} classified these ``apparitions'' according to the date of the discovery of the emission line.
Hereafter, we keep the name apparition proposed by \citet{preston11} since the second and third emissions are not always observed at each pulsation
cycle and consequently are like  ``ghost emissions''.

\textbf{The first apparition} is a strong blueshifted \ha emission in RR\,Lyrae stars that was first reported by \cite{struve48b}.
Because it was the first such detection, \cite{preston11} called this strongest \ha emission ``first apparition''.
It occurs immediately before the luminosity maximum, around pulsation phase $\varphi=0.92$, and may be caused by the sum of
five shocks that are theoretically expected according to \cite{fokin97}; the strongest of these shocks is initiated by the $\kappa$-mechanism.
The first self-consistent theory was proposed by \cite{fadey04}.
The model considers a stationary shock wave within an homogeneous medium consisting solely of atomic hydrogen with the five first
bound levels and the continuum, allowing a computation of the hydrogen emission line profiles.
Their most remarkable result is that the velocity inferred from Doppler shifts of the Balmer emission lines is roughly one-third of the shock wave velocity,
with a mean value of about 130\kmsn, depending on the pulsation cycle.

\textbf{The second apparition} was discovered by \citet{gillet88} as a small blueshifted emission component within the large \ha absorption profile in two stars: X\,Ari (a non-Blazhko RR\,Lyrae star), and RR\,Lyr itself.
This small emission appears during the bump, that is, before the luminosity minimum near the pulsation phase $\varphi=0.72$.
The bump was interpreted by \citet{hill72} as the consequence of a shock or collision between atmospheric layers.
\citet{gillet88} attributed  this ``second apparition'' to a secondary shock that is produced when the outer atmospheric layers, following a ballistic motion, impact inner layers.
For RR\,Lyr, this secondary emission seems absent at the Blazhko minimum phase ($\psi=0.50$), but it is observable immediately before the Blazhko maximum ($\psi=0.97$).
This means that an amplitude threshold may exist to induce this secondary shock.

\textbf{The third apparition} was recently found in some RR\,Lyrae stars \cite{preston11}, but not in RR\,Lyr itself.
This weak and redshifted emission shoulder appears within the \ha absorption near the pulsation phase $\varphi=0.30$, after maximum light, and occurs in Blazhko and non-Blazhko stars.
It has never been detected before in RR\,Lyr.
\citet{preston11} did not explain its physical origin, but \citet{chadid13} suggested that this emission component might be caused by a weakly supersonic
and infalling shock wave at the beginning of atmospheric compression.
However, it should be noted that these authors did not explain how such a compression shock alone can produce redshifted emission.
In contrast to the explanation by \citet{chadid13}, \citet{gillet14} proposed that the redshifted emission component of \ha might
be produced by the main shock.
Emission this weak should only be observed when the main shock propagates far enough from the photosphere, that is, when the shock intensity is very high.
In this case, the emission profile would become a P-Cygni type profile, and would be the consequence of the large extension of the expanding atmosphere.

\medskip
In this paper, we present for the first time the observation of the third \ha emission in RR\,Lyr itself.
In Sect.\,2 we describe observational and data reduction processes.
The detection of the hydrogen line third apparition is presented in Sect.\,3, while
its physical origin is discussed in Sect.\,4.
Finally, some concluding remarks are given in Sect.\,5.

%=====================================================================================%
\section{Observations and data analysis}
%---------------------------------------------------------------------------------%
\subsection{Data acquisition}

Since we are especially interested in the \ha third emission that occurs in the pulsation phase interval $0.2\lesssim \varphi \lesssim0.4$, we reexamined previous published spectra and obtained new detections of the third emission at this particular step of the pulsation and for various Blazhko phases.
We finally used data obtained from spectrographs between 1997 and 2017.

\begin{itemize}
\item \textsc{elodie} spectrograph: Attached to the 193\,cm telescope at the Observatoire de Haute-Provence, it is described in \citet{1996A&AS..119..373B}.
Observations were performed in $1996-1997$ in the context of a survey led by D. Gillet and published in  \cite{1999A&A...352..201C}.
We used some spectra from these observations to verify the third emission during different Blazhko phases.
Typical exposure times were between $5-8$\,min, leading to a
signal-to-noise ratio (S/N) of about 50 per pixel for a resolving power of $R = 42\,000$. It appears that this relatively low S/N is not enough for a suitable detection of the third emission, and we are left with only two Blazkho phases.
In addition, for the night of 1997-08-30, four spectra were stacked to improve the S/N per pixel to 80.
This echelle spectrograph allows observing the whole visible domain (3\,900-6\,800\,\AA), but it suffers from overlapping in blue orders that prevents us from observing some important lines such as the \ion{Na}{i} doublet or the Balmer H$_{\beta}$ line.

\smallskip
\item \textsc{aurelie} spectrograph: Attached to the 152\,cm telescope at the Observatoire de Haute-Provence, it is described in \citet{gillet94}. Data used in this paper were collected during two separate nights in September 2013 (16 spectra) and September 2014 (12 spectra). High-resolution spectra ($R=22\,700$) were obtained with a typical exposure time of 10\,min, leading to a mean S/N of about 85 per pixel. The spectral domain is relatively short: 6\,500-6\,700\,\AA.

\smallskip
\item \textsc{eShel} spectrograph: Attached to an automated 35\,cm telescope at the Observatoire de Chelles (France), the fiber-fed \textsc{eShel} spectrograph was described in \citet{2011IAUS..272..282T} and was built by Shelyak Instruments\footnote{\url{http://www.shelyak.com}}. A few spectra were gathered from December 2016 to April 2017 to search for the third emission at other Blazhko phases.
The exposure times varied between 10 and 30\,min (less than 4\% of the pulsation period), providing an S/N per pixel of between 50 and 100 for a resolving power of $R=11\,000$. The spectral domain is spread over $4\,300-7\,100$\,\AA. The detector used for the \textsc{eShel} spectrograph is an Atik 460EX CCD (Sony ICX694 sensor).
\end{itemize}

Observations are summarized in Table \ref{longobslog}, where the columns provide the date of the night, the corresponding
Julian Date, the telescope (Tel.) and observatory (Obs.), the attached spectrograph (Spectro.), its resolving power, its resolution element RE, the spectral domain, the typical exposure time T$_{\textrm{exp}}$
used for the night, the mean S/N per pixel in the $\lambda$6630 region, the number $N$ of spectra considered, and finally the start and end of the pulsation phase and the Blazhko phase for each night.

%-- Version 3 : 20170829
{\setlength{\tabcolsep}{4.5pt}
\begin{table*}
\centering
\caption{Characteristics of the RR\,Lyr spectra.}
\label{longobslog}
\begin{tabular}{cccccccccccccc}
\hline
Night & JD                           & Tel. & Obs. & Spectro.   & Resolving & RE &$\lambda_{\rm start} - \lambda_{\rm end}$ & T$_{\textrm{exp}}$ & S/N & N & $\varphi_{start}$ & $\varphi_{end}$ & $\psi$ \\
{\scriptsize (yyyy-mm-dd)} & {\scriptsize (-2\,400\,000)} &    &     &           & power     & (pixel) &       (\AA)                                 & (min)              &     &   &                   &                 &        \\
\hline
1997-08-09                 & 50670                        & 193
 & 
OHP & \textsc{elodie}   & 42\,000 & 2.1 &  $3\,800-6\,800$                          &  8                 &  70 & 1 &  0.311            & 0.321           & 0.61   \\
1997-08-30                 & 50691                        & 193
 &  OHP & \textsc{elodie}   & 42\,000 & 2.1 &  $3\,800-6\,800$                          &  5                 &  98 & 4 &  0.277            & 0.301           & 0.14   \\
2013-09-04                 & 56540                        & 152
 & 
OHP &\textsc{aurelie}   & 22\,700 & 2.8 & $6\,500-6\,700$                           & 10                 &  80 &16 &  0.074            & 0.289           & 0.95   \\
2014-09-14                 & 56915                        & 152
 &  OHP &\textsc{aurelie}   & 22\,700 & 2.8 & $6\,500-6\,700$                           & 10                 &  90 &12 &  0.332            & 0.471           & 0.86   \\
2016-12-03                 & 57726                        & 35
 &  CHELLES & \textsc{eShel} & 11\,000 & 3.2 & $4\,300-7\,100$                           & 30                 & 114 & 2 &  0.207            & 0.281           & 0.54   \\
2017-03-26                 & 57838                        & 35
 & 
CHELLES & \textsc{eShel} & 11\,000 & 3.2 & $4\,300-7\,100$                           & 30                 &  98 & 1 &  0.227            & 0.264           & 0.44   \\
2017-03-29                 & 57842                        & 35
 &  CHELLES & \textsc{eShel} & 11\,000 & 3.2 & $4\,300-7\,100$                           & 10                 &  52 & 2 &  0.231            & 0.256           & 0.54   \\
2017-04-02                 & 57846                        & 35
 & 
CHELLES & \textsc{eShel} & 11\,000 & 3.2 & $4\,300-7\,100$                           & 10                 &  70 & 4 &  0.255            & 0.304           & 0.64   \\
\hline
\end{tabular}
\end{table*}
}

%-----------------------------------------------------------------------%
\subsection{Data reduction}

All observations were reduced using classical operations such as preprocessing (bias and dark subtraction, flat-fielding, masking of bad pixels,
spectrum extraction, wavelength calibration, correction for instrumental response), and the observations provide spectra in the heliocentric rest frame.
For \textsc{elodie} spectra, the online pipeline was used as described in \citet{chadid96}.
The \textsc{aurelie} and \textsc{eShel} observations were reduced using subpackages of the the Audela\footnote{\url{http://www.audela.org}} software, SpcAudace\footnote{\url{http://spcaudace.free.fr}} and a dedicated echelle package, respectively, that were originally implemented.

Then we normalized the spectra to the local continuum in the \ha region, and a Savitsky-Golay filter \citep{sg67} was applied for visual detection to remove noise while preserving the spectral resolution of each data set. Nevertheless, all measures were made on unfiltered spectra.
% This filter makes it possible, better than a Gaussian filter, to get rid of the high frequency noise while preserving weak bumps, such as that of the third emission. Nevertheless, all measures were done on unfiltered spectra.}

Wavelengths were further corrected for the heliocentric velocity and radial velocity of RR\,Lyr, which is also called the $\gamma$-velocity, using $RV=-73.5$\,\kms \citep{chadid96}.
Our \ha line profile observations show that the absorption line center is not centered on the laboratory wavelength (represented as a vertical line in Figs.\,\ref{20130904multi}-\ref{1997-2017}), unlike the line profiles presented in \citet{preston11}, which seem to present anomalous wavelength corrections.

Finally, Figs.\,\ref{20130904multi}-\ref{20170326NaD} represent the spectra in the detector histogram mode. Depending on the spectrograph, the resolution element is between 2.1 and 6.0 pixels (see Table \ref{longobslog}). Thus, the resolving power is directly visible in the spectra.

%---------------------------------------------------------------------------------%
\subsection{Ephemerides computation}

Periods and phases both change for the pulsation and also for the Blazhko variations. Since we are
interested in peculiar phases, it is fundamental to use a current and adapted ephemeris.
Recently, \citet{leborgne14} showed that the pulsation period alternates between two primary pulsation states, defined as pulsation over a
a ``long''  (0.56684\,d) and a ``short''  (0.56682\,d) period,
with intervals of 13-16\,yr.
A direct consequence is that the ephemeris used by \cite{1999A&A...352..201C} for the \textsc{elodie} observations (1997) is not relevant since it concerns the epoch
1982-1989.
We therefore preferably used ephemerides provided by \citet{leborgne14} for the \textsc{elodie} and \textsc{aurelie} observations.
For the more recent 2016-2017 data, an epoch out of that for the ephemeris
%{I think there is some confusion here - the
%plural of "ephemeris" is "ephemerides". Mostly I can tell when
%you meant to use the plural by the verb form you use, but please
%check carefully throughout for instances like here where it could
%be either singular or plural; also for the subheading here} 
provided by \citet{leborgne14}, the pulsation period $P_P$ was computed using the difference between
two $O-C$ minima from the GEOS RR\,Lyr web database\footnote{\url{http://rr-lyr.irap.omp.eu}} and close to our 2016-2017 observations. The reference HJD maximum light ($HJD0_P$) was also provided by this database.
A summary of the ephemerides for the different data sets is provided in Table\,\ref{ephemeris}.

{\setlength{\tabcolsep}{4.4pt} % Au lieu de 6pt
\begin{table}[H] %[!h]
\caption{Ephemerides used to compute the pulsation and Blazhko phase.}
\centering
\label{ephemeris}
\begin{tabular}{c|cc|cc}
\hline
Epoch & \multicolumn{2}{c}{Pulsation} & \multicolumn{2}{|c}{Blazhko} \\
\hline
      & $HJD0_P$                    & Period                      & $HJD0_B$   & Period \\
      & {\scriptsize (-2\,400\,000)}& $P_P$ (d) & {\scriptsize (-2\,400\,000)} & $P_B$ (d)  \\
\hline
1997  & 50\,456.7090\tablefootmark{a} & 0.5668174\tablefootmark{b} & 49\,631.312\tablefootmark{c} & 39.06\tablefootmark{b} \\
2013  & 56\,539.3428\tablefootmark{a} & 0.5667975\tablefootmark{b} & 56\,464.481\tablefootmark{d} & 39.0\tablefootmark{b} \\
2014  & 56\,914.5507\tablefootmark{a} & 0.56684\tablefootmark{b}   & 56\,881.627\tablefootmark{d} & 39.0\tablefootmark{b} \\
2017  & 57\,861.6319\tablefootmark{a} & 0.566793\tablefootmark{a}  & 57\,354.322\tablefootmark{d} & 39.0\tablefootmark{b} \\
\hline
\end{tabular}
\tablefoot{Maximum light HJD and periods are from\\
\tablefoottext{a}{GEOS} \\
\tablefoottext{b}{\citet{leborgne14}} \\
\tablefootmark{c}{\citet{1999A&A...352..201C}} \\
\tablefootmark{d}{This paper}
}
\end{table}
}

It is more difficult to determine the Blazhko phase because the Blazhko period $P_B$ does not follow the variation of the pulsation period, but presents rather erratic changes \citep{leborgne14}.
Moreover, during 2014, the photometric $O-C$ were historically low (close to nil), preventing an easy computing of the Blazhko period.
Therefore, in order to determine the maximum light amplitude dates $HJD0_B$, we developed an innovative method based on equivalent width and main shock velocities based on our three years of spectral observations, which will be described in a forthcoming paper (Gillet et al 2018).
We expect an uncertainty of $\pm 2$\,days on $HJD0_B$ and of $\pm 0.2$\,day on $P_B$.
The ephemerides were computed with the period $P_{B}=39.0$~d \citep{leborgne14}.
However, for the 1997 observations, the maximum light amplitude date was taken from \cite{1999A&A...352..201C} and the Blazhko period is given by \citet{leborgne14}, that is, $P_{B\,1997}=39.06$~d.
Blazhko ephemerides used for the different data sets is summarised in Table\,\ref{ephemeris} too.

%=======================================================================================%
\section{Observation of the hydrogen third emission in \rrlyn}

%-----------------------------------------------------------------------%
\subsection{Evidence of the third emission}
%--- Description methode de sosutraction des ajustements :

\begin{figure}%[!h]
  \centering
  \includegraphics[width=0.97\hsize]{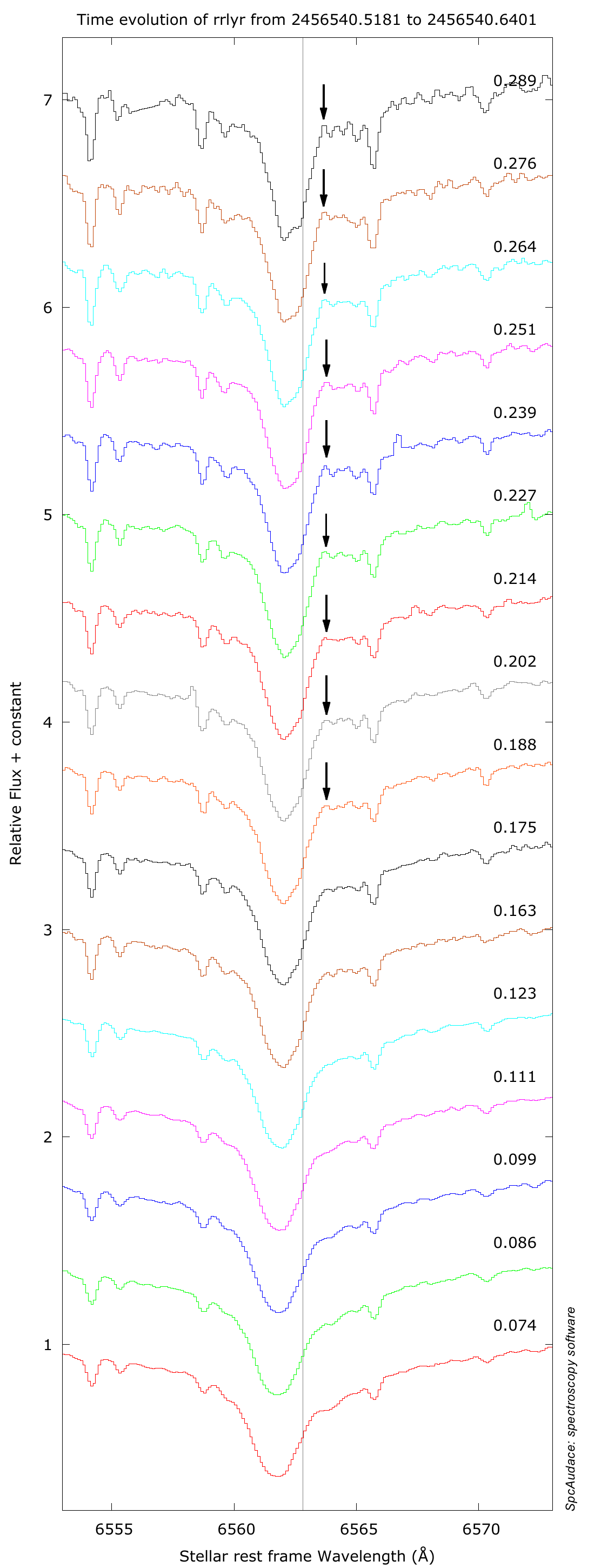}
  \caption{Evolution of the \ha line profile of \rrly for the night 2013-09-04 (Blazhko phase $\psi=0.95$).
The third emission is the weak hump indicated by the arrows, and it occurs for $\varphi=0.188-0.289$.
Weak absorption features are telluric lines.
The vertical line indicates the \ha line laboratory wavelength in the stellar rest frame.}
  \label{20130904multi}
\end{figure}

\begin{figure}%[!h]
  \centering
  % height=9.15in ; 11.6 gnuplot
  \includegraphics[width=0.97\hsize]{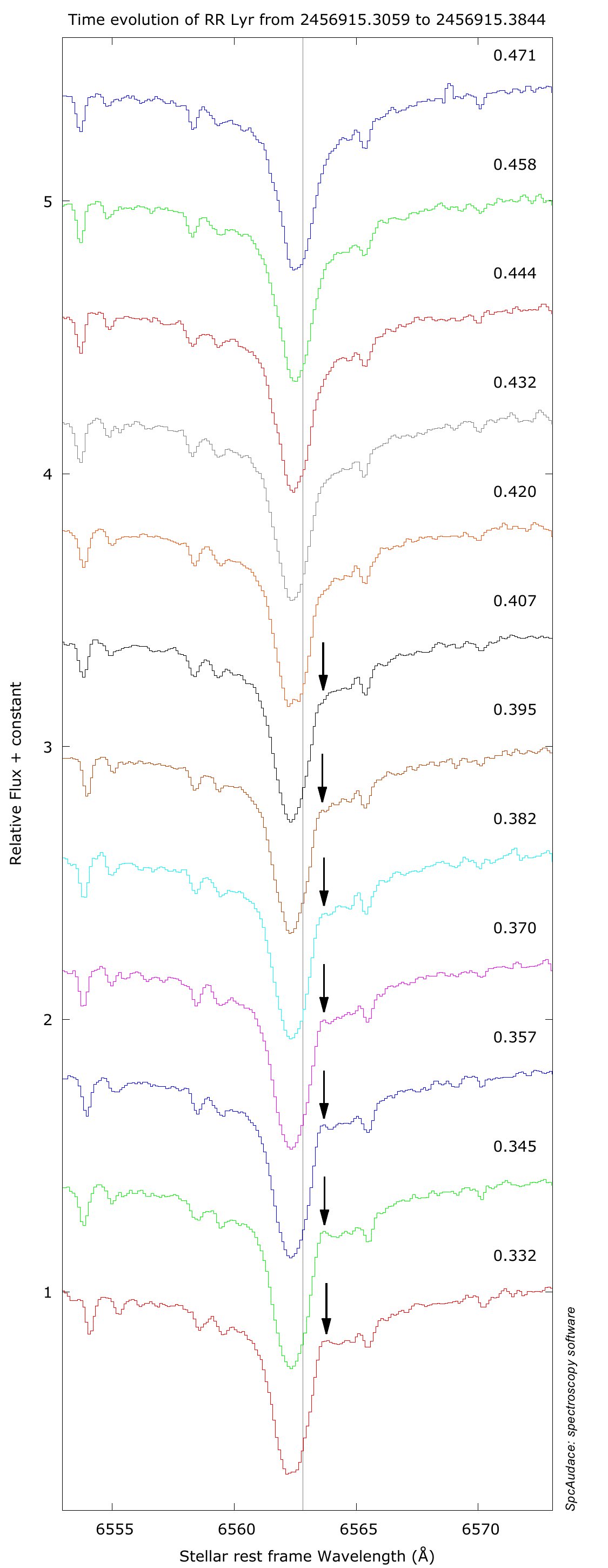}
  \caption{Evolution of the \ha line profile of \rrly for the night 2014-09-14 (Blazhko phase $\psi=0.86$).
The third emission is the weak hump indicated by the arrows, and it occurs for $\varphi=0.332-0.407$.
The weak absorption features are telluric lines.
The vertical line indicates the \ha line laboratory wavelength in the stellar rest frame.}
  \label{20140914multi}
\end{figure}

Figures\,\ref{20130904multi} (2013-09-04) and \ref{20140914multi} (2014-09-14) show the evolution of the \ha line profile during the phase intervals $\varphi=0.074-0.289$ and $\varphi=0.332-0.471,$ respectively.

No appreciable structure appears in the blue wing of the \ha absorption profile, while the red wing is clearly affected by a weak hump for
the relevant phase intervals $\varphi=0.188-0.289$ and $\varphi=0.332-0.407$.
This hump represents the so-called \ha third emission.
We note that the deeper analysis was conducted on the 2013 and 2014 observations, since their temporal sampling is best suited for the follow-up of the third emission.
We emphasize that all the considered observations of Table\,\ref{longobslog} also present this third emission, regardless of the phase $\psi$ of the Blazhko cycle.

%---------------------------------------------------------------------------%
\subsection{Extraction of the third emission}

To highlight the third emission within the \ha profile, which manifests itself as a hump in the spectra, we chose to subtract the different absorption profiles in this spectral region.
Since the atmospheric dynamics may strongly affect the stellar line profile, we did not use a synthetic stellar spectrum since these latter are computed for a static atmosphere. The different shock waves occurring in the atmosphere during the pulsation cycle (five main shock waves for RR Lyr, see \citet{fokin97}) induce high velocity gradients and a strong rise (by up to two or three times) in the level of turbulence.

We used individual (Gaussian) profiles instead, regardless of their origin, stellar or telluric, using the PySpecKit library \citep{pyspeckit11}.
This Python package uses as input the number of lines present in the spectral domain considered together with their wavelengths, which are fixed, and it provides the residual flux and widths for each component.
The \ha absorption profile, which presents Stark wings, was modeled using three Gaussians of different widths.
Other features have a telluric origin ($\lambda=6557.21$, 6558.12, 6560.43, 6562.68, 6562.82, 6563.56, and 6564.26\,\AA).
We note that the computations were made in the geocentric rest frame to match the telluric lines, and then we transformed back to the stellar rest frame.
We emphasize that no known lines (whether stellar or of telluric origin) are present at the third emission expected wavelength.

\begin{figure}[H] %[!ht]
  \centering
  \includegraphics[width=\hsize]{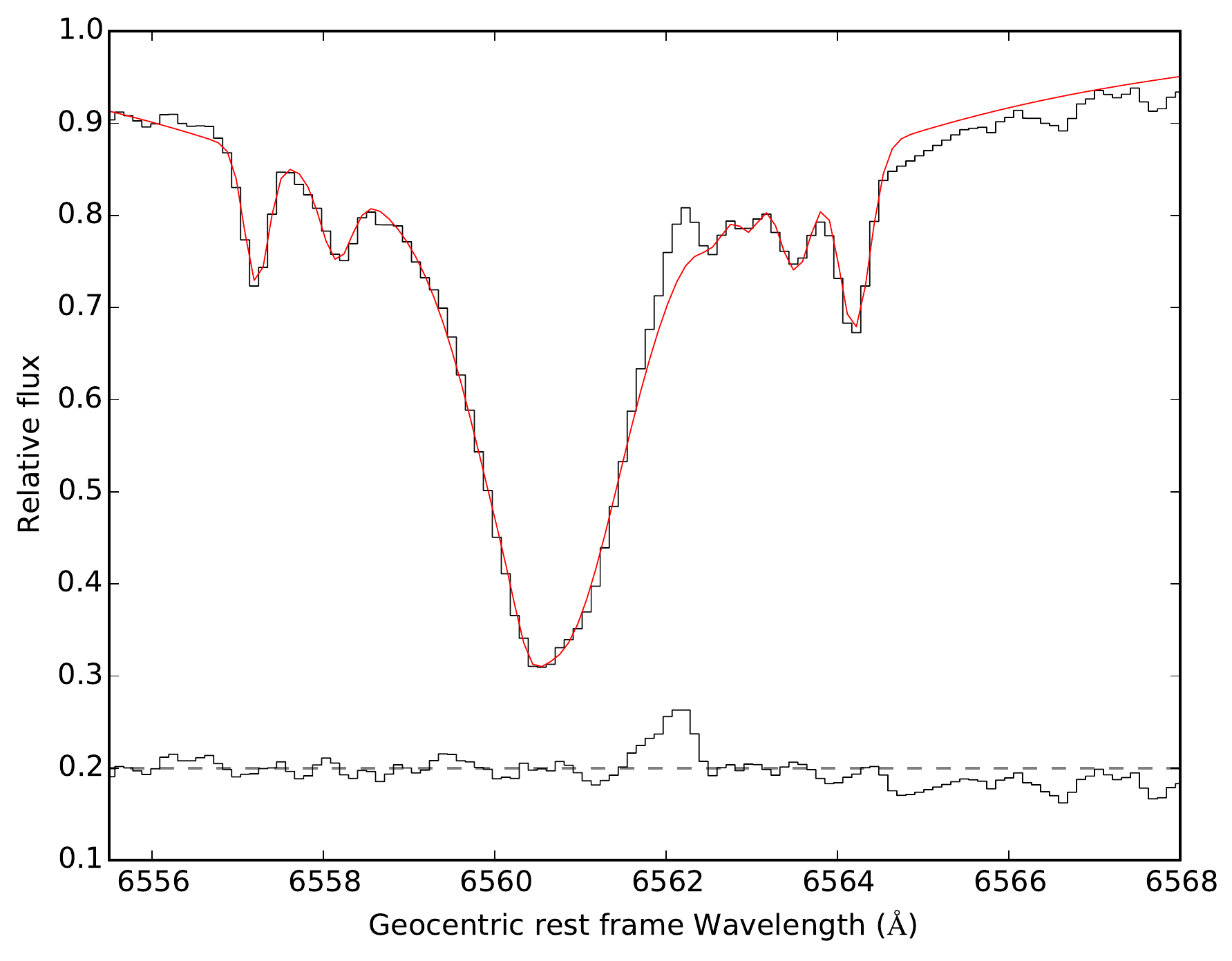}
  \caption{Observed \ha spectral domain (black) and its resulting fit (red) for the \ha absorption profile and seven telluric lines computed for the
2013-09-04 night at $\varphi=0.239$.
The difference is plotted at the bottom of the figure with a $+0.2$ vertical offset for clarity.
The residual clearly shows the third emission component within the red wing of the \ha absorption profile.}
  \label{fittingexample}
\end{figure}

\begin{figure}%[!h]
  \centering
  \includegraphics[width=\hsize]{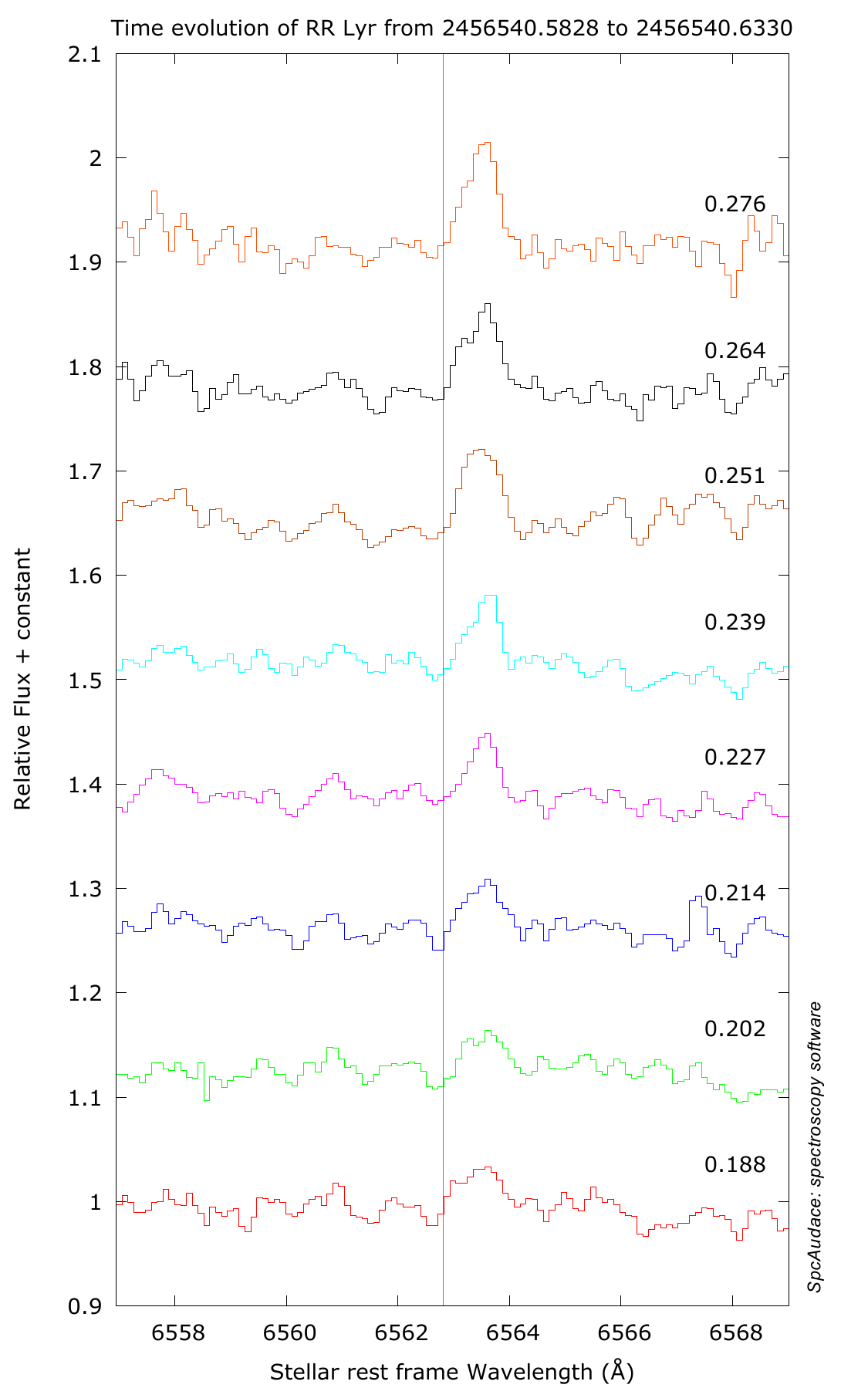}
  \caption{Evolution of the residuals for the night 2013-09-04 for pulsation phases $\varphi=0.188-0.276$.
The vertical line indicates the \ha line laboratory wavelength in the stellar rest frame.}
  \label{20130904over183-234}
\end{figure}

\begin{figure}%[!h]
  \centering
  \includegraphics[width=\hsize]{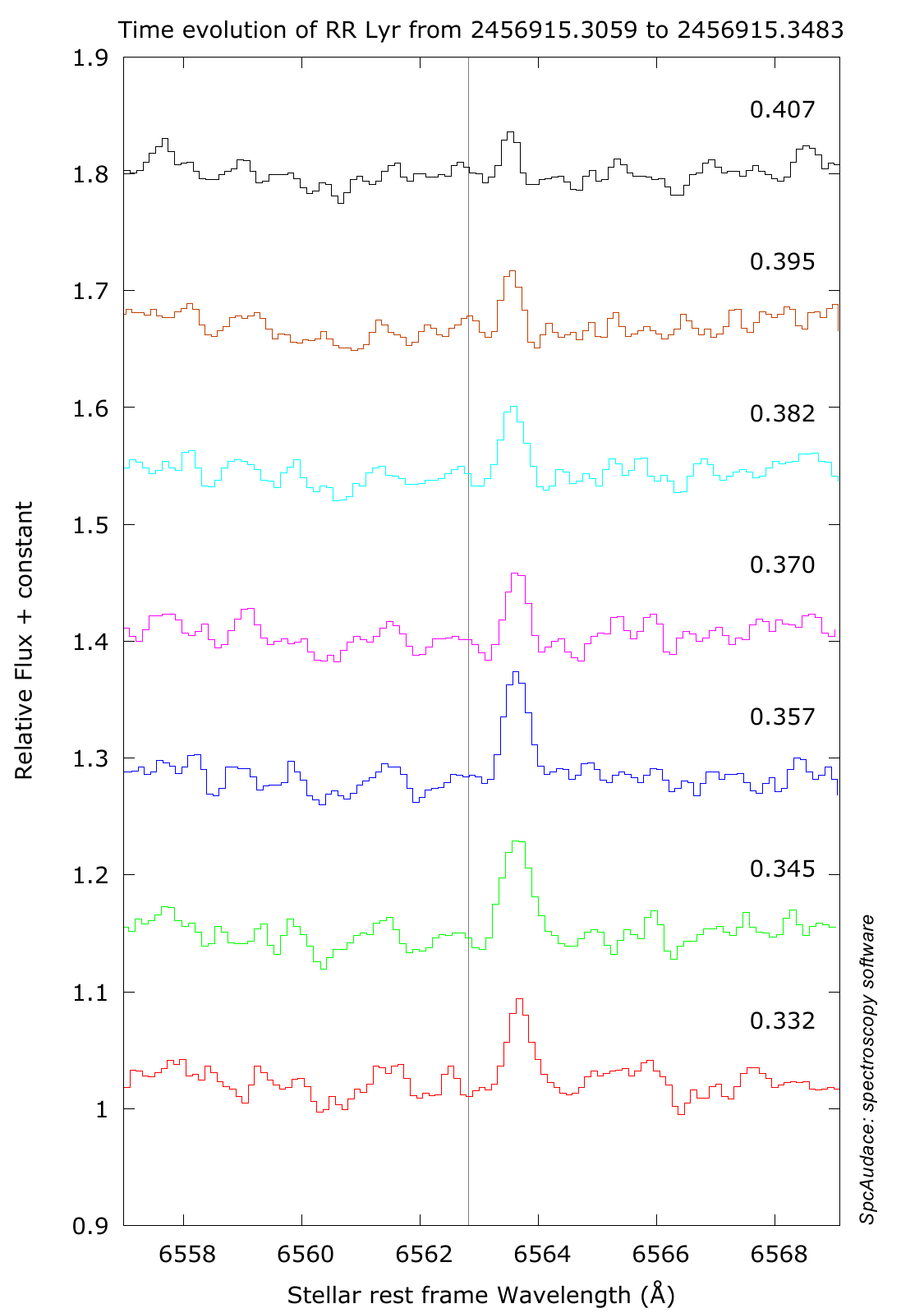}
  \caption{Evolution of the residuals for the night 2014-09-14 for pulsation phases $\varphi=0.332-0.407$.
The vertical line indicates the \ha line laboratory wavelength in the stellar rest frame.}
  \label{20140914over299-362}
\end{figure}

An example of the resulting fit is provided in Fig.\ref{fittingexample} together with the residual obtained by subtracting
the fit from the initial profile: the third emission around 6562\,\AA\ (geocentric rest frame) is now clearly visible.
This procedure was carried out for all the spectra used in this paper, and examples of residuals at different phases are presented
in Figs.\,\ref{20130904over183-234} and \ref{20140914over299-362}.

%-- Fig {fitting_exemple.pdf} :  Fig. 2    exemple fitting

%-- Fig {20130904over183-234} :  Fig. 6   soustraction gaussiennes 2013

%-- Fig {20140914over299-362} :  Fig. 5    soustraction gaussiennes 2014

%---------------------------------------------------------------------------------%
\subsection{Spectral characteristics of the third emission}

\begin{figure}%[!h]
  \centering
  \includegraphics[width=\hsize]{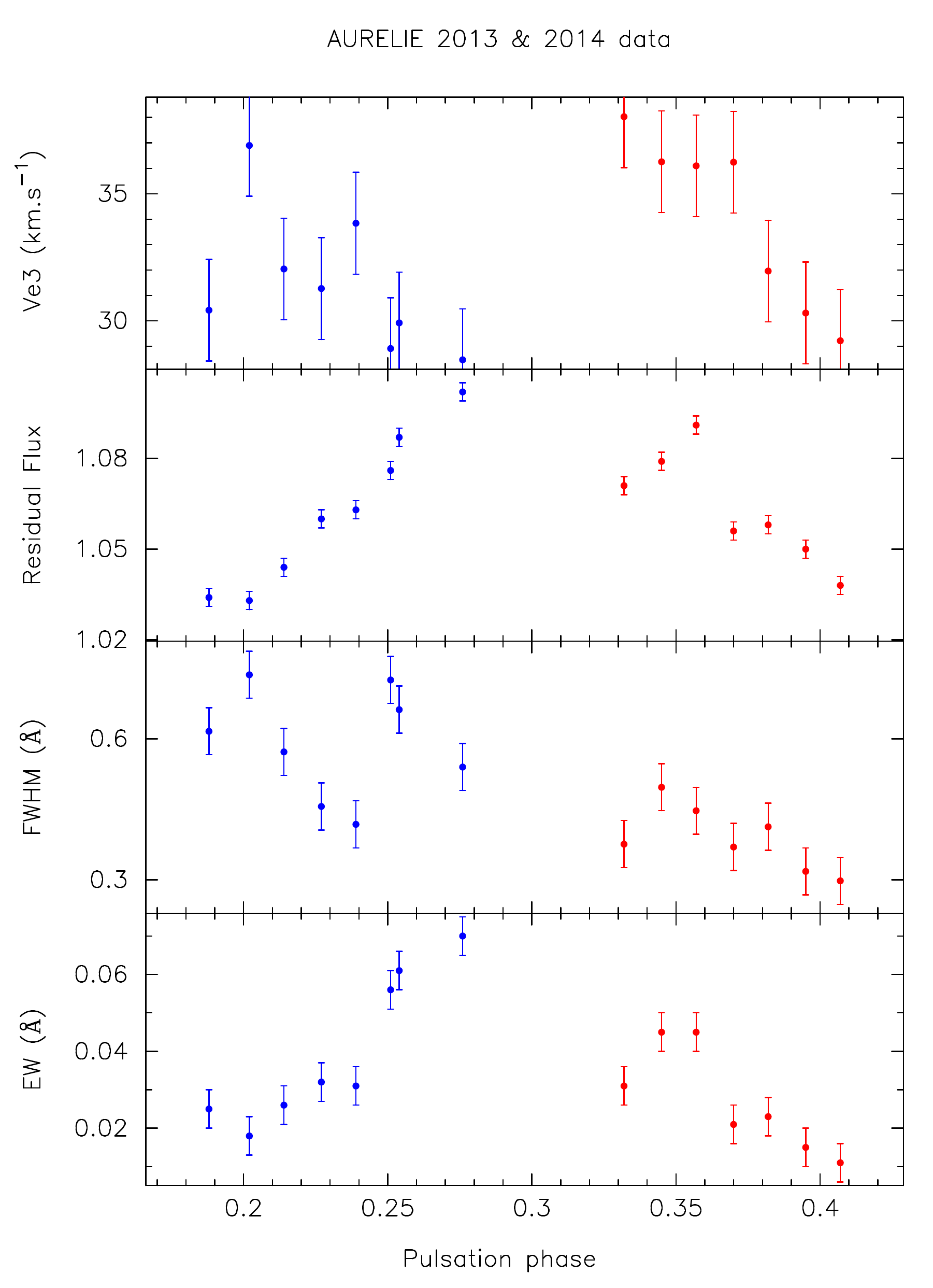}
  \caption{Third-emission line characteristics for the nights 2013-09-04 (blue) and 2014-09-14 (red).
From top to bottom we show the Doppler velocity in the stellar frame, the RF, the
FWHM, and the EW. }
  \label{ve3}
\end{figure}

Hence, the fitting process allows a direct study of the third emission profile, and some characteristics of the line can be derived, such as the associated velocity (hereafter $V_{e3}$), the full width at half-maximum (FWHM), the residual flux (RF), and the equivalent width (EW), as a function of the pulsation phase.
For the two \textsc{aurelie} nights, these quantities are represented in Fig.\,\ref{ve3}.

%-- Fig {plot_Ve3.pdf} :  Evolution de Ve3

The typical uncertainties are about 2\,\kms for the velocity data, 0.5\,\% for the flux, and
80\,m\AA\ and 8\,m\AA\ for the FWHM and EW quantities, respectively.
The RF peaks at about 10\,\% above the continuum, at a level similar to other measurements in other RR\,Lyrae stars.
While the RF and EW show a continuous variation between the two epochs, the FWHM
instead shows a constant value for the 2013 data and a decreasing behavior during 2014.

For the two nights, the velocity seems to follow a regular slow-down at a rate of about 200\,\cmsd for the 2014 data
and about 100\,\cmsd for the 2013 data, which shows that a breaking mechanism is active.
The difference between the slope associated with each night might be due either to a different atmospheric
dynamics or to a different Blazhko phase, since the nights are separated by about 10\% of the Blazhko period.

The variation in the FWHM and EW suggests that temperature effects are maximum at about $\varphi \sim 0.3$, and they may be linked to the inward ballistic motion.

%---------------------------------------------------------------------------%
\subsection{Evolution of the third emission}
%-- Autres observations de la 3e emission et relation avec l'effet Blazhko : -----------------%

\begin{figure}%[!h]
  \centering
  \includegraphics[width=\hsize]{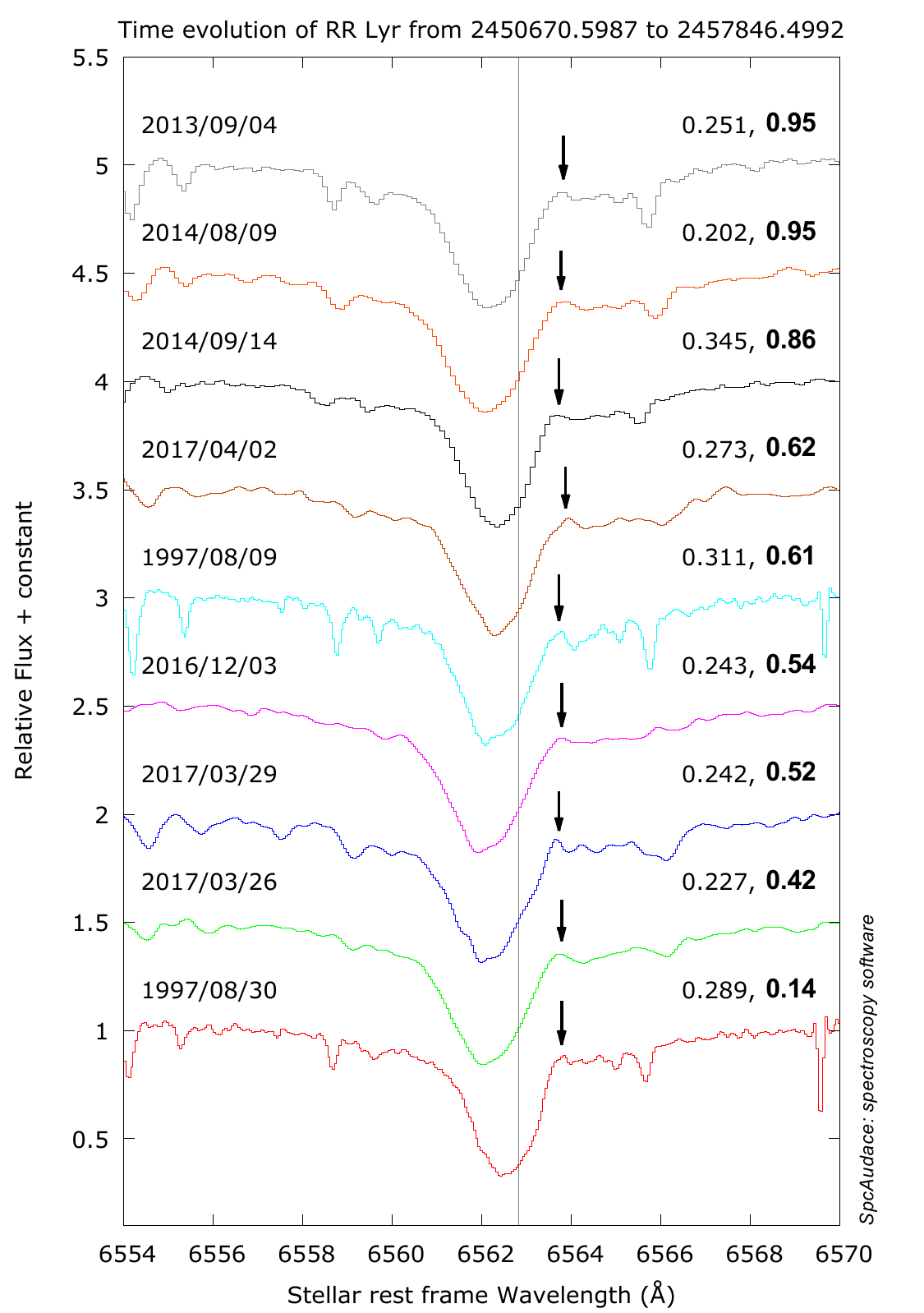}
  \caption{\ha profile of \rrly for the 1997 to 2017 observations. Pulsation and Blazhko (in bold) phases (right) and observation dates (left) are indicated. The third emission is present at all Blazhko phases despite the different spectral resolutions and S/N. The detector pixel is shown in each spectrum. The vertical line shows the \ha line laboratory wavelength in the stellar rest frame.}
  \label{1997-2017}
\end{figure}

Since the atmospheric dynamics seems to change between the 2013 and 2014 observations, it is interesting to extend the study of these spectral characteristics to other Blazhko phases.
Figure\,\ref{1997-2017} presents spectra obtained over the 20 years of observations presented in Table\,\ref{longobslog}.
It is clear that the third apparition is observed at all Blazhko phases.
While the flux of the third emission varies by up to 10\% during the pulsation phase, there is no such a difference
with the Blazhko phase: the third emission seems not larger during the two Blazhko maxima observed in 2013 and 2014 ($\psi=0.95$) than during the Blazhko minima of 2016 and 2017 ($\psi=0.54$).

%-- Fig {1997-2017} :  Fig. 8

To summarize, the third emission is present for about 20\% of the pulsation period, from $\varphi=0.188-0.407$, with a typical flux
of about 10\% of the continuum.
This emission is clearly redshifted, by about 35\kms in the stellar rest frame.
Finally, its characteristics seems to be independent of the Blazhko phase.

%=======================================================================================%
\section{Physical origin of the hydrogen third emission}
%---------------------------------------------------------------------------------%
\subsection{Still-active infalling motion}

\begin{figure}%[!h]
  \centering
  \includegraphics[width=\hsize]{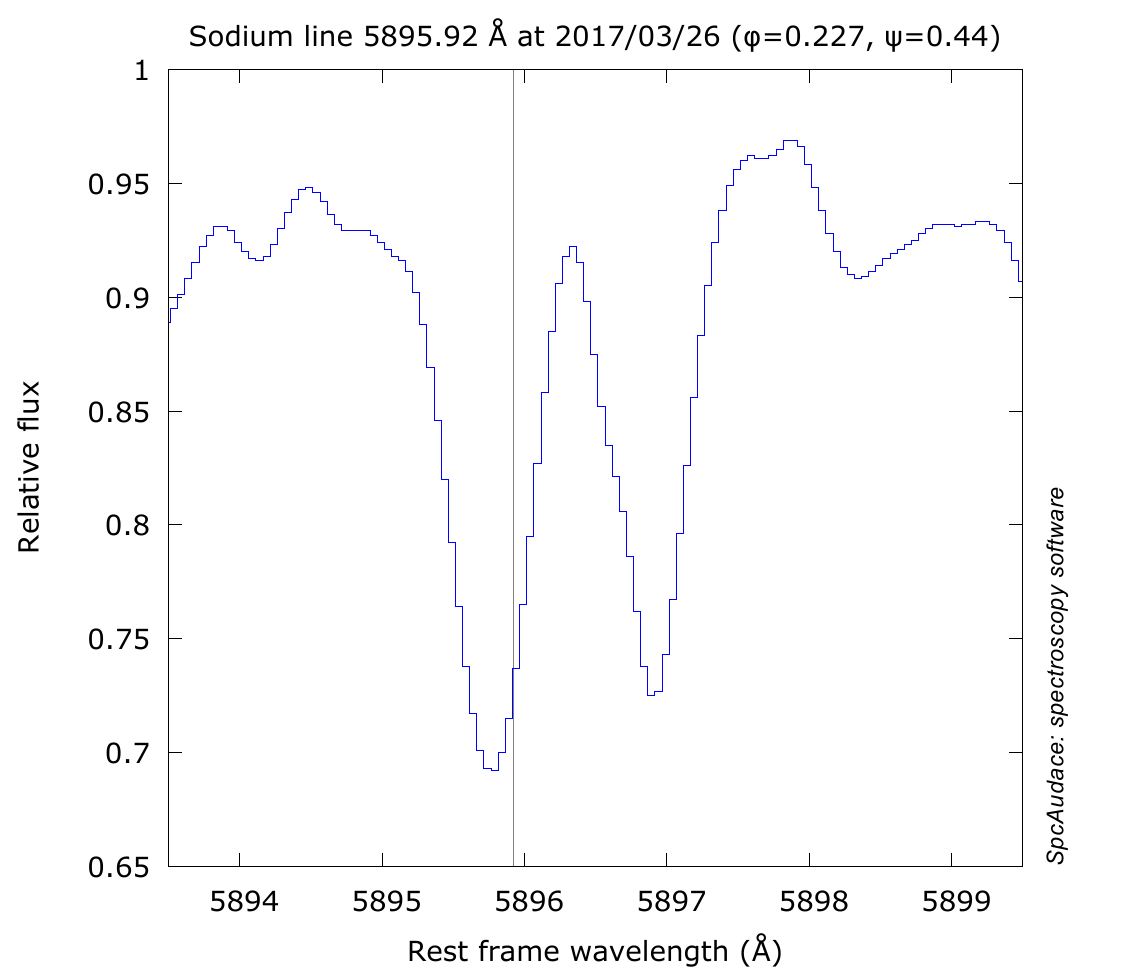}
  \caption{\ion{Na}{i} $\lambda$5895.92 (D1) profiles of RR\,Lyr for $\varphi=0.227$ on the night of 2017-03-26.
The Blazhko phase is $\psi=0.44,$ that is, close to the minimum.
The vertical line indicates the sodium line laboratory wavelength in the stellar rest frame.
Note that two strong telluric lines are blanketing the center of the D2 component, which is different to the D1 component \citep{hobbs78}.}
  \label{20170326NaD}
\end{figure}

According to the observations reported in this paper, the third hydrogen emission first appears near the pulsation phase $\varphi=0.188$
(Fig.\,\ref{20130904multi}) and is observed until $\varphi=0.407$ (Fig.\,\ref{20140914multi}).
At this stage, photospheric layers should rise within the atmosphere since the shock wave passage starts at $\varphi=0.92$.
However, the upper atmosphere may still follow an infall ballistic motion.
Only observable in the \textsc{eShel} observations because of
the poor order overlapping in the \textsc{elodie} data, the \ion{Na}{i} line profile doubles during night 2017-03-26 ($\varphi \sim 0.227$, $\psi=0.44$, Fig.\,\ref{20170326NaD}), which is representative of the other \textsc{eShel} spectra as well.

While the blue component is related to the outward shock wave passage, the red component refers to the material falling back on the star.
Therefore, at $\varphi \sim 0.227$, there is still a free-fall motion of the upper atmosphere.
Unfortunately, since only one spectrum (which in addition has T$_{\textrm{exp}} = 30$\,min) has been obtained in that night, it is not possible
to follow the evolution of the two components.
Thus, we can infer from this observation that during the third-emission phase, the shock wave is located between the two \ion{Na}{i} layers,
and it is related to the line doubling phenomenon.

%-----------------------------------------------------------------------%
\subsection{Supersonic infalling motion}

The redshifted velocity of the \ha third emission gives an idea of the compression rate in the atmosphere.
For the 2013 observations with pulsation phases $\varphi=0.188-0.276$, the velocity of the redshifted emission remains
approximately constant with a value of about 31\kmsn, while for
the 2014 observations with pulsation phases $\varphi=0.332-0.407$, the velocity continuously decreases from 39\kms to 32\kms (Fig.\,\ref{ve3}) around an average value of 35\kmsn.
If the pulsation cycles are different, it seems clear that the velocity of the redshifted emission component maintains an approximately similar average value with pulsation phase.
Thus, the atmospheric compression rate does not vary much in the phase interval $\varphi=0.188-0.276$, while it seems to slowly increase in the interval $\varphi=0.332-0.407$.
There would therefore be a first phase during which the compression ratio would remain stable before starting to slowly decrease.

In 2014, the intensity of the emission reached a maximum near the pulsation phase $\varphi\sim 0.350$ (see Fig. \ref{20140914over299-362}),
which must correspond to the strongest atmospheric heating.
In 2013, the strongest emission intensity must certainly occur after $\varphi=0.276,$ but unfortunately, we do not have observations after this phase.
On the other hand, in 2013, the third emission appears to be wider than in 2014.
This would indicate that the temperature, and therefore the compression rate, would be higher during this pulsation cycle.
However, it is clear that new observations of good quality are necessary to confirm this type of variation from one pulsation cycle to another.

The two sodium absorption components are separated by about 60\kmsn, corresponding to the velocity difference between the two layers in opposite motion.
The upward velocity of the gas is maximum immediately behind the shock front.
However, because of the strong gravity of RR\,Lyr ($g_\star\sim250$\cmsd at the RR\,Lyr photospheric radius, see \citealp{fossati14}), this velocity
decreases continuously when moving away from the shock front, in the wake.
Because the red component is shifted by about 50\kms with respect to the sodium laboratory wavelength in the stellar rest frame, this means
that the infalling motion of the highest part of the atmosphere is largely supersonic (Mach number around 5 for a sound speed near 10\kmsn).
Consequently, the shock propagates in a higher-density medium compared to a static atmosphere with a classical decreasing density law.
Future observations of the evolution of the \ion{Na}{i} blue component velocity should test this assumption of a deceleration with the pulsation phase.

%-----------------------------------------------------------------------%
\subsection{Shock front velocity}

No direct observations are available, such as a blue component emission on the spectra during the phase interval $\varphi=0.188-0.407$, that could help to directly measure the shock wave velocity.

According to \citet{gillet14}, the shock front velocity varies from 130\kms at $\varphi=0.902$ to 115\kms at $\varphi=0.927$.
These values are based on the shock models developed by \citet{fadey04}, who derived

\begin{equation}
V_{\textrm{shock}} \cong 3c\frac{(\lambda_{e1}-\lambda_{0})}{\lambda_{0}}
\label{vshock1}
,\end{equation}

\noindent
where $\lambda_{e1}$ is the wavelength of the maximum intensity of the H$\alpha$ first-apparition emission and $\lambda_{0}$ its laboratory
wavelength.
On the other hand, because the D3 helium near the pulsation phase $\varphi=1.04$ is a P-Cygni profile, it implies that the shock velocity
corresponds to the expansion rate of the gas shell, which is
about 60\kms \citep{gillet14}.
Thus, the shock strongly decelerates between the pulsation phases 0.902 and 1.04.
Since in the phase interval $\varphi=0.188-0.407$ the shock does not produce any emission component within the \ha profile (see below), we can
reasonably estimate that the shock intensity is weak and that the velocity of the front is much lower than 60\kmsn.

Moreover, since the blue component of the sodium line is weakly blueshifted (-10\kmsn), while the red component is strongly redshifted (50\kmsn),
it can even be considered that the shock may have a velocity lower than that of the infalling gas.
Ultimately, in the phase interval $\varphi=0.188-0.407$, the shock wave is no longer radiative, hence no emission component appears.
It may even be close to its complete dissipation in the upper atmosphere.
A stellar shell of a thickness of a few hundred kilometers and a radius on the order of 1.35 times the photospherical radius is probably dynamically unstable if its velocity is not sufficiently high (supersonic).

%-----------------------------------------------------------------------%
\subsection{Weakening of the radiative shock shell}

When the intensity of the main shock wave is strong enough, a significant extension of the atmosphere, and a consecutive P-Cygni profile, are possible.
This was clearly observed at $\varphi=1.04$ with the D3 helium line \citep{gillet16}.
Since it is reasonable to assume a continuation of the propagation of the shock shell at least until $\varphi=1.04$, \citet{gillet14} suggested
that the third \ha emission component is also a P-Cygni profile.
This explanation is plausible even though the emission is not centered on the laboratory wavelength in the stellar rest frame.
As shown by \citet{WBB83}, an advancing shell can produce a redshifted emission depending on the spectral line formation parameters such as the
local line width, the geometrical extension of the emitting layer, the optical thickness, and especially the velocity gradient.

However, our recent observation on 2017-03-26 (see Fig.\,\ref{20170326NaD}) of the sodium line doubling at $\varphi \sim 0.227$ shows that the infalling
motion of the upper layers of the atmosphere compresses the low layers that are located immediately above the photosphere.
The plasma created by this intense compression is most likely the cause of the third \ha emission component.
Moreover, the shock wave velocity is lower than 60\kmsn, as discussed above.
Consequently, the explanation by \citet{WBB83} is not necessary or is not valid here to justify the third emission because the shock velocity is probably too weak to induce pertinent parameters for a redshifted third emission.

%-----------------------------------------------------------------------%
\subsection{Atmospheric structure near $\varphi=0.3$}

At the phase $\varphi\sim0.3$, the deepest photospheric layers are close to their maximum radius \citep{fokin97}.
Their velocity is therefore close to zero, and they will begin to fall back onto the star.
At the same time, the main shock is located in the upper atmosphere at approximately 1.35 times the photospheric radius.
It always propagates outward, but at low velocity since it does no longer produce a blueshifted emission component within the \ha profile: the shock is no longer radiative.

In the meantime, the layers above the shock have not yet completed their infalling motion because of the sodium line doubling.
Their average velocity is clearly supersonic since the red component is shifted by about 50\kmsn.

The supersonic infalling motion of the highest atmospheric layers compresses the gas located in front of the shock more and more.
The latter also contributes to the compression process by its opposite movement.
Therefore, the third \ha emission is produced within the strongly compressed zone, hence high-temperature zone, that exists immediately above the shock front.

%-----------------------------------------------------------------------%
\subsection{Compression zone}

With the extraction technique used (a sum of Gaussians), the consecutive profile of the third emission appears rough. In particular, the shape of the line profile does not seem accurate enough (reality of asymmetries, width variability, etc.) to derive a physical interpretation, especially in 2013.
However, in particular through the EW and the RF, it is clear that the intensity of the emission increases after its appearance, reaches a maximum, and then decreases before finally disappearing (see Fig.\,\ref{ve3}).

This behavior is also expected to be observed for the FWHM, but this is not the case for the 2013 night.
This night shows deeper telluric lines and a lower S/N than the 2014 night.
Thus, for 2013, it is more difficult to extract the third emission.

The limit of the extraction method could also affect the $V_{e3}$ measured radial velocities.
Considering the 2013 night, the $V_{e3}$ decreases from +36\kms to +30\kms ($\varphi \sim0.20$ and $0.25$), while for the 2014 night, the decrease is from +38\kms to +29\kms ($\varphi \in [0.3;0.4]$).
This general trend, a low, continuous decrease (starting from $\varphi \sim 0.2$ for the 2013 night) for both nights, is related to an increasing atmospheric compression.
Therefore, the compression zone producing the emission is slowed down during its fall onto the star.
Compared to the 2013 night, the $V_{e3}$ value for the 2014 night is higher (+38\kmsn at $\varphi=0.33$).
This would confirm that from one Blazhko cycle to another (which differ by about 10\%), the amplitudes of the atmospheric dynamics can be very different.

This point seems to be well correlated with the observation of shock velocities that vary significantly from one cycle to another (Gillet et al 2018).
Unfortunately, to confirm this point for these two nights of 2013 and 2014, we have no observations around phase 0.92 with
which to determine an estimate of the shock velocity.

%-----------------------------------------------------------------------%
\subsection{Comparison of the second and third apparitions}

The second apparition consists of a small blueshifted emission, while the third apparition is a very weak redshifted emission.
Both are interpreted as a consequence of the compression of the highest atmospheric layers on the deepest layers.

The small blueshifted emission (second apparition) was discovered by \cite{gillet88} near the pulsation phase $\varphi=0.72$.
This emission is visible during a relatively small phase interval ($\Delta\varphi\approx0.1$).
Its physical origin could be the result of a collision between the layers of the upper atmosphere with the photospheric layers during
the infall phase of the ballistic motion.
During this phase interval, all atmospheric layers contract and a luminosity bump occurs in the light curve \citep{gillet88}.
The emission occurs just before the minimum radius at $\varphi\sim0.92$, when the main shock emerges from the photosphere \citep{fokin97}.
Because the gravity of RR\,Lyr is much higher ($g_\star\sim250$\cmsdn) than that of Classical Cepheids
($g_\star\sim20-110$\cmsdn), W\,Virginis stars ($g_\star\sim15$\cmsdn), or RV\,Tauri stars ($g_\star\sim4$\cmsdn), we must expect a high atmospheric
compression rate at the end of the ballistic motion of the atmosphere.

Consequently, depending on the amplitude of the ballistic motion, that is, of the shock intensity of the preceding pulsation cycle, the compression
rate may be sufficiently high to produce a strong heating of the gas, and a consecutive luminosity bump is observed.
Because the infall velocity of the gas can become supersonic, especially at the end of the ballistic motion, a shock front moving outward in mass
when the pulsation phase increases can be produced.
This is the reason for the appearance of a blueshifted emission.
The observation of RR\,Lyr presented in \cite{gillet88} shows that the emission is blueshifted by 79\kmsn, which is indeed a supersonic speed.
Finally, this emission is induced by the ballistic motion of atmospheric layers that is induced by, and thus its amplitude depends on, the main shock emerging in the photosphere at the beginning of the cycle.
Therefore, the second emission is not directly produced by the main shock because it is not yet present in the atmosphere.

The very small redshifted emission (third apparition) occurs around the pulsation phase $\varphi=0.30$ during an appreciable phase
interval ($\Delta\varphi\sim0.2$), twice as long as that of the second apparition.
When the third emission appears near the phase $\varphi\sim0.19$, the deepest photospheric layers propagate outward in the atmosphere while
the highest layers have not completed their infalling motion (sodium line doubling).
Consequently, the motion of low and high atmospheric layers originates in two different but consecutive pulsation cycles,
introducing a coupling between these pulsation cycles.
This coupling is absent when all atmospheric layers have enough
time to complete their dynamical relaxation, as for low-amplitude pulsators.

Finally, the third apparition is the direct consequence of the high stellar gravity of RR\,Lyr.
It generates a strong infall motion of the highest part of the atmosphere.
This may explain why the third emission is not observed in other radially pulsating stars, except perhaps for long-period Cepheids, 
which have a stellar gravity ten times lower than that of RR Lyrae, and which is is interpreted as a P-Cygni profile 
\citep{2008A&A...489.1263N, 2014A&A...568A..72G}. 
This profile would be the consequence of the shock shell propagation in the high atmosphere.

The physical origin of the secondary and third emissions is thus not the same.
The second emission is a consequence of the differential infall velocity between high and deep layers.
The infall motion of the deepest photospheric layers is slowed down by the increasing compression on the high-density subphotospheric layers,
while the highest layers continue their accelerated infall motion because of the low gas density in the high atmosphere.
The third emission is the encounter between two layers of opposite motion: the lowest and highest parts of the atmosphere.
This causes a compression zone that is located in front of the shock wave, which induces the observed redshifted emission.
Thus, there are two different driving mechanisms: (i) the lowest layers are driven by the main shock, which rises in the atmosphere and reverses
the movement of the layers falling onto the star; (ii) the layers are driven by gravity.

Finally, the two emissions are the result of a compression phenomenon caused by atmospheric dynamics.
They are not simply a consequence of the passage of a rising shock wave within the atmosphere.

%-----------------------------------------------------------------------%
\subsection{Influence of the Blazhko phase}

In Fig.\,\ref{1997-2017} we show the third apparition in the \ha line profile independent of the Blazhko phases.
In first approximation, the third emission does not present a higher intensity at Blazhko maxima than at other Blazhko phases. This is not at all the case when the \ha blueshifted emission occurs when the main shock passes through the photosphere near $\varphi=0.94$ \citep{1997A&A...319..154C}.

The third emission appears when the shock is located in the high atmosphere ($\varphi=0.188-0.407$).
In this phase interval, the shock is already far away from the photosphere ($1.35\,R_{ph}$), and as discussed above, the shock velocity becomes very weak.
It is even probable that the shock loses its radiative nature, that is to say, it is no longer able to produce hydrogen emission.
We recall that the third \ha emission component is produced by the atmospheric compression located in front of the shock.

Consequently, the intensity of the third \ha emission does not depend solely on the shock velocity of the previous pulsation cycle, but more directly on the atmospheric dynamics.
Thus, the impact of the Blazhko phase, that is, the shock amplitude, is significantly reduced at this distance from the star.
The intensity of the third emission also depends on the intensity of the main shock of the current pulsation cycle since it is
located between the ascending and descending atmospheric layers.
The intensity of the third emission is a function of the intensities of the two main waves: that of the current cycle and that of the preceding cycle.
It should be noted, however, that the influence of this latter shock is not as direct as the influence of the first shock, but only because of the ballistic motion of the atmosphere it has initiated.

Finally, the intensity of the third emission depends more or less directly on the intensity of the two main shock waves that
belong to two consecutive pulsation cycles because their intensity is considerably reduced at high altitude.
This may be the reason why the intensity dependence of the third \ha emission is not so obvious.

New observations with high S/N should enable us to quantify this marginal dependence of the Blazhko phase on the strength of the third emission.

%=======================================================================================%
\section{Conclusion}

We presented for the first time a set of observations of the \ha third emission in RR\,Lyr, the prototype of the RR\,Lyrae stars.
This very weak emission appears as a hump within the red wing of the \ha line.
It is observed around $\varphi\sim0.30$ at a stellar rest frame velocity of about 30\kmsn.
This phenomenon lasts for about 20\% of the pulsation period during pulsation phases $\varphi=0.188-0.407$.
The maximum intensity of this third emission with respect to the continuum is 13\%.

For the night 2017-03-26 ($\varphi \sim 0.227$, $\psi=0.44$), when the \ha third emission was present, the \ion{Na}{i} lines double.
This line doubling is intepreted as opposite motions between the upper atmosphere that is still falling back onto the star, and deeper layers dragged by the main outward shock wave.
The high stellar gravity of RR\,Lyr plays a decisive role in the development of the shock by attenuating the amplitude of the ballistic motion of the atmosphere.
The strong damping of the shock intensity prevents the production of any emission component within the \ha profile that would be formed in the shock wake.
Thus, the shock is no longer radiative, and is probably close to its complete dissipation in the upper layers of the atmosphere.

Conversely, the free fall of the upper atmosphere is highly supersonic because the sodium red component is redshifted by 50\kmsn.
The \ha third emission would be due to the excitation of hydrogen atoms by the dynamical compression induced by the infalling motion of the atmosphere.
From measuring the velocity of the redshifted emission, it appears that the compression region is supersonic as well.

It would be interesting to determine whether this compression process in the upper atmosphere also occurs in other RR Lyrae stars and even in other types of pulsating stars such as RV Tauri stars and classical Cepheids, since these stars also have high-intensity
shocks and significant atmospheric extensions.

Finally, the data and work presented in this paper demonstrate further the increasing role of the amateur spectroscopy community in stellar surveys.

%---------------------------------------------------------------------------------%
%-- Acknowledgements :
\bigskip

\begin{acknowledgements}
We thank \textit{Lux Stellarum} and the French OHP-CNRS/PYTHEAS for their support.
The present study has used the SIMBAD data base operated at the Centre de Donn\'ees Astronomiques (Strasbourg, France) 
and the GEOS RR\,Lyr data base hosted by IRAP (OMP-UPS, Toulouse, France), created by J.F. Le Borgne.
This research made use of SpcAudace, an open-source spectroscopic toolkit hosted at \url{http://spcaudace.free.fr} and written by B. Mauclaire (ARAS group, France).
We also thank P. Valvin for his useful reading. 
We especially thank the referee for their very careful reading of the manuscript and pertinent remarks.
And we gratefully acknowledge Astrid Peter for her very wise reading of the final version of this paper.
\end{acknowledgements}

%= Bibliography ===========================================================%

\bibliographystyle{astroads}
\bibliography{rrlyr_biblio}

\begin{thebibliography}{30}
\expandafter\ifx\csname natexlab\endcsname\relax\def\natexlab#1{#1}\fi
\expandafter\ifx\csname href\endcsname\relax
  \def\href#1#2{}\fi
\expandafter\ifx\csname urllinklabel\endcsname\relax
  \def\urllinklabel{[LINK]}\fi
\expandafter\ifx\csname adsurllinklabel\endcsname\relax
  \def\adsurllinklabel{[ADS]}\fi

\bibitem[{{Baranne} {et~al.}(1996){Baranne}, {Queloz}, {Mayor}, \&
  al.}]{1996A&AS..119..373B}
{Baranne}, A., {Queloz}, D., {Mayor}, \& al. 1996, \aaps, 119, 373
 \href{http://adsabs.harvard.edu/abs/1996A%26AS..119..373B}{\adsurllinklabel}

\bibitem[{{Bla{\v z}ko}(1907)}]{bla1907}
{Bla{\v z}ko}, S. 1907, Astronomische Nachrichten, 175, 325
 \href{http://adsabs.harvard.edu/abs/1907AN....175..325B}{\adsurllinklabel}

\bibitem[{{Chadid} \& {Gillet}(1996)}]{chadid96}
{Chadid}, M. \& {Gillet}, D. 1996, \aap, 308, 481
 \href{http://adsabs.harvard.edu/abs/1996A%26A...308..481C}{\adsurllinklabel}

\bibitem[{{Chadid} \& {Gillet}(1997)}]{1997A&A...319..154C}
---. 1997, \aap, 319, 154
 \href{http://adsabs.harvard.edu/abs/1997A%26A...319..154C}{\adsurllinklabel}

\bibitem[{{Chadid} {et~al.}(1999){Chadid}, {Kolenberg}, {Aerts}, \&
  {Gillet}}]{1999A&A...352..201C}
{Chadid}, M., {Kolenberg}, K., {Aerts}, C., \& {Gillet}, D. 1999, \aap, 352,
  201
 \href{http://adsabs.harvard.edu/abs/1999A%26A...352..201C}{\adsurllinklabel}

\bibitem[{{Chadid} \& {Preston}(2013)}]{chadid13}
{Chadid}, M. \& {Preston}, G.~W. 2013, \mnras, 434, 552
 \href{http://adsabs.harvard.edu/abs/2013MNRAS.434..552C}{\adsurllinklabel}

\bibitem[{{Fadeyev} \& {Gillet}(2004)}]{fadey04}
{Fadeyev}, Y.~A. \& {Gillet}, D. 2004, \aap, 420, 423
 \href{http://adsabs.harvard.edu/abs/2004A%26A...420..423F}{\adsurllinklabel}

\bibitem[{{Fokin} \& {Gillet}(1997)}]{fokin97}
{Fokin}, A.~B. \& {Gillet}, D. 1997, \aap, 325, 1013
 \href{http://adsabs.harvard.edu/abs/1997A%26A...325.1013F}{\adsurllinklabel}

\bibitem[{{Fossati} {et~al.}(2014){Fossati}, {Kolenberg}, {Shulyak}, \&
  al.}]{fossati14}
{Fossati}, L., {Kolenberg}, K., {Shulyak}, D.~V., \& al. 2014, \mnras, 445,
  4094
 \href{http://adsabs.harvard.edu/abs/2014MNRAS.445.4094F}{\adsurllinklabel}

\bibitem[{{Gillet}(2014)}]{2014A&A...568A..72G}
{Gillet}, D. 2014, \aap, 568, A72
 \href{http://adsabs.harvard.edu/abs/2014A%26A...568A..72G}{\adsurllinklabel}

\bibitem[{{Gillet} {et~al.}(1994){Gillet}, {Burnage}, {Kohler}, \&
  al.}]{gillet94}
{Gillet}, D., {Burnage}, R., {Kohler}, D., \& al. 1994, \aaps, 108, 181
 \href{http://adsabs.harvard.edu/abs/1994A%26AS..108..181G}{\adsurllinklabel}

\bibitem[{{Gillet} \& {Crowe}(1988)}]{gillet88}
{Gillet}, D. \& {Crowe}, R.~A. 1988, \aap, 199, 242
 \href{http://adsabs.harvard.edu/abs/1988A%26A...199..242G}{\adsurllinklabel}

\bibitem[{{Gillet} \& {Fokin}(2014)}]{gillet14}
{Gillet}, D. \& {Fokin}, A.~B. 2014, \aap, 565, A73
 \href{http://adsabs.harvard.edu/abs/2014A%26A...565A..73G}{\adsurllinklabel}

\bibitem[{{Gillet} {et~al.}(2016){Gillet}, {Sefyani}, {Benhida}, \&
  al.}]{gillet16}
{Gillet}, D., {Sefyani}, F.~L., {Benhida}, A., \& al. 2016, \aap, 587, A134
 \href{http://adsabs.harvard.edu/abs/2016A%26A...587A.134G}{\adsurllinklabel}

\bibitem[{{Ginsburg} \& {Mirocha}(2011)}]{pyspeckit11}
{Ginsburg}, A. \& {Mirocha}, J. 2011, {PySpecKit: Python Spectroscopic
  Toolkit}, Astrophysics Source Code Library
 \href{http://adsabs.harvard.edu/abs/2011ascl.soft09001G}{\adsurllinklabel}

\bibitem[{{Hill}(1972)}]{hill72}
{Hill}, S.~J. 1972, \apj, 178, 793
 \href{http://adsabs.harvard.edu/abs/1972ApJ...178..793H}{\adsurllinklabel}

\bibitem[{{Hobbs}(1978)}]{hobbs78}
{Hobbs}, L.~M. 1978, \apj, 222, 491
 \href{http://adsabs.harvard.edu/abs/1978ApJ...222..491H}{\adsurllinklabel}

\bibitem[{{Kov{\'a}cs}(2016)}]{kovacs16}
{Kov{\'a}cs}, G. 2016, Coms. of the Konkoly Observatory Hungary, 105, 61
 \href{http://adsabs.harvard.edu/abs/2016CoKon.105...61K}{\adsurllinklabel}

\bibitem[{{Le Borgne} {et~al.}(2014){Le Borgne}, {Poretti}, {Klotz}, \&
  al.}]{leborgne14}
{Le Borgne}, J.~F., {Poretti}, E., {Klotz}, A., \& al. 2014, \mnras, 441, 1435
 \href{http://adsabs.harvard.edu/abs/2014MNRAS.441.1435L}{\adsurllinklabel}

\bibitem[{{Nardetto} {et~al.}(2008){Nardetto}, {Groh}, {Kraus}, {Millour}, \&
  {Gillet}}]{2008A&A...489.1263N}
{Nardetto}, N., {Groh}, J.~H., {Kraus}, S., {Millour}, F., \& {Gillet}, D.
  2008, \aap, 489, 1263
 \href{http://adsabs.harvard.edu/abs/2008A%26A...489.1263N}{\adsurllinklabel}

\bibitem[{{Pickering} {et~al.}(1901){Pickering}, {Colson}, {Fleming}, \&
  al.}]{pick1901}
{Pickering}, E.~C., {Colson}, H.~R., {Fleming}, W., \& al. 1901, \apj, 13
 \href{http://adsabs.harvard.edu/abs/1901ApJ....13..226P}{\adsurllinklabel}

\bibitem[{{Preston}(2011)}]{preston11}
{Preston}, G.~W. 2011, \aj, 141, 6
 \href{http://adsabs.harvard.edu/abs/2011AJ....141....6P}{\adsurllinklabel}

\bibitem[{{Preston} {et~al.}(1965){Preston}, {Smak}, \&
  {Paczynski}}]{preston65}
{Preston}, G.~W., {Smak}, J., \& {Paczynski}, B. 1965, \apjs, 12, 99
 \href{http://adsabs.harvard.edu/abs/1965ApJS...12...99P}{\adsurllinklabel}

\bibitem[{{Savitzky} \& {Golay}(1964)}]{sg67}
{Savitzky}, A. \& {Golay}, M. J.~E. 1964, Anal. Chem., 36, 1627
 \href{http://dx.doi.org/10.1021/ac60214a047}{\urllinklabel}

\bibitem[{{Schwarzschild}(1952)}]{schwa52}
{Schwarzschild}, M. 1952, {Transactions of the IAU \textsc{VIII}}, Vol.~8
  (Roma, {Oosterhoff}, P. Th. (CUP)), 811


\bibitem[{{Smolec}(2016)}]{smolec13}
{Smolec}, R. in , 2016, 37th Meeting of the Polish Astronomical Society, ed.
  A.~{R{\'o}{\.z}a{\'n}ska}M.~{Bejger}, Vol.~3, 22--25
 \href{http://adsabs.harvard.edu/abs/2016pas..conf...22S}{\adsurllinklabel}

\bibitem[{{Struve} \& {Blaauw}(1948)}]{struve48b}
{Struve}, O. \& {Blaauw}, A. 1948, \apj, 108, 60
 \href{http://adsabs.harvard.edu/abs/1948ApJ...108...60S}{\adsurllinklabel}

\bibitem[{{Thizy} \& {Cochard}(2011)}]{2011IAUS..272..282T}
{Thizy}, O. \& {Cochard}, F. Active OB Stars: Structure, Evolution, Mass Loss,
  and Critical Limits, ed. , C.~{Neiner, }G.~{Wade, }G.~{Meynet} \&
  G.~{Peters}, 282--283
 \href{http://adsabs.harvard.edu/abs/2011IAUS..272..282T}{\adsurllinklabel}

\bibitem[{{Wagenblast} {et~al.}(1983){Wagenblast}, {Bertout}, \&
  {Bastian}}]{WBB83}
{Wagenblast}, R., {Bertout}, C., \& {Bastian}, U. 1983, \aap, 120, 6
 \href{http://adsabs.harvard.edu/abs/1983A%26A...120....6W}{\adsurllinklabel}

\bibitem[{{Xiong} {et~al.}(1998){Xiong}, {Cheng}, \& {Deng}}]{XCD98}
{Xiong}, D.~R., {Cheng}, Q.~L., \& {Deng}, L. 1998, \apj, 500, 449
 \href{http://adsabs.harvard.edu/abs/1998ApJ...500..449X}{\adsurllinklabel}

\end{thebibliography}

%*************************************************************%
\end{document}